\pdfoutput=1
\documentclass[reprint,amsmath,amssymb,aps,]{revtex4-1}

\usepackage{graphicx}% Include figure files
\usepackage{dcolumn}% Align table columns on decimal point
\usepackage{bm}% bold math
\usepackage{physics}
\usepackage{color}
\usepackage{hyperref}% add hypertext capabilities
\begin{document}

\preprint{APS/123-QED}

\title{Shapes of fluid membranes with chiral edges}% Force line breaks with \\

\author{Lijie Ding$^1$}
    \email{lijie\_ding@brown.edu}\textbf{}
\author{Robert A. Pelcovits${^1}{^2}$}
\author{Thomas R. Powers${^1}{^3}$}

\affiliation{$^1$Department of Physics, Brown University, 182 Hope Street, Providence, RI 02912 USA
}
\affiliation{$^2$Brown Theoretical Physics Center and Department of Physics, Brown University, 182 Hope Street, Providence, RI 02912 USA
}
\affiliation{$^3$School of Engineering and Department of Physics, Brown University, 182 Hope Street, Providence, RI 02912 USA}

\date{\today}% It is always \today, today,

\begin{abstract}
    We carry out Monte Carlo simulations of a colloidal fluid membrane composed of chiral rod-like viruses. The membrane is modeled by a triangular mesh of beads connected by bonds in which the bonds and beads are free to move at each Monte Carlo step. Since the constituent viruses  are experimentally observed to twist only near the membrane edge, we use an effective energy that favors a particular sign of the geodesic torsion of the edge.  The effective energy also includes membrane bending stiffness, edge bending stiffness, and edge tension. We find three classes of membrane shapes resulting from the competition of the various terms in the free energy: branched shapes, chiral disks, and vesicles. Increasing the edge bending stiffness smooths the membrane edge, leading to correlations among the membrane normal at different points along the edge. We also consider membrane shapes under an external force by fixing the distance between two ends of the membrane, and find the shape for increasing values of the distance between the two ends. As the distance increases, the membrane twists into a ribbon, with the force eventually reaching a plateau.
\end{abstract}

%Use showkeys class option if keyword
\maketitle

%\tableofcontents

\section{Introduction}
    Fluid membranes are ubiquitous in biological systems and exhibit various shapes due to their fluidity and the constraints of fixed area and fixed volume. While closed membrane vesicles show a wide range of shapes including pears, discocytes, stomatocytes, and toroids~\cite{seifert1997configurations}, membranes with free edges can form shapes other than flat disks.For example, colloidal membranes composed of aligned rod-like chiral viruses  in the presence of a polymer depletant are typically found to have open edges and form twisted ribbon shapes. The handedness of a ribbon, defined by the handedness of helical edge of the ribbon, is determined by the intrinsic chirality of the viruses; reversing the chirality of the viruses reverses the handedness of the ribbons~\cite{barry2010entropy,gibaud2012reconfigurable,zakhary2014imprintable}. Lipid bilayer membranes with free edges also play a role during the formation of vesicles~\cite{boal1992topology,chernomordik2008mechanics,huang2017formation}, and can be stabilized by reducing the line tension of the edge~\cite{fromherz1983lipid,zhao2005monte}. Likewise, liposomes exposed to increasing levels of the protein talin form lipsomes with stable holes, cup-shaped liposomes, and, finally, lipid bilayer sheets~\cite{saitoh1998,TuOuYang2003}. Helical ribbons are also seen as intermediate states in the formation of self-assembled tubes from lipid molecules~\cite{Barclay_etal2014,SelingerSpectorSchnur2001}.

    In this paper, we use Monte Carlo simulations to study the configurations that arise in a simple effective model for colloidal membranes. A theoretical model of the mechanics of colloidal membranes must account for the bending energy of the membrane, the chiral liquid crystal energy associated with the orientational ordering of the rodlike colloidal particles, and finally the energy associated with the free edges of the membrane. Models that have been developed to date include phenomenological Landau models \cite{kaplan2010theory,tu2013theory,tu2013instability,jia2017chiral}, entropically-motivated models \cite{kang2016entropic,gibaud2017achiral} and hard body simulations \cite{yang2012self, xie2016interaction}.

    There are two competing effects governing the alignment of the rodlike viruses in a colloidal membrane. One is the tendency for the rods to line up side by side. The other is a tendency for the rods to twist due to their intrinsic chirality. These two tendencies are incompatible and thus the twist is confined to a region near the membrane edge. The thickness of this region is known as the twist penetration depth \cite{deGennes1972}). If the twist penetration depth is small compared to the lateral dimensions of the membrane, as is often the case in these colloidal membranes, the  liquid crystalline degrees of freedom can be accounted for by an effective theory in which the local degrees of freedom do not appear explicitly, and the energy depends only on geometric properties of the surface. This approach was taken by Jia et al.~\cite{jia2017chiral} who accounted for the liquid crystalline degrees of freedom with an effective edge energy which includes an edge tension term involving the length of the perimeter, a bending energy cost for the curvature of the edge, and a chiral term involving the geodesic torsion of the edge. The geodesic torsion is the rate that the normal to the surface twists around the edge of the surface~\cite{kleman}. Even when using this simplified model it is difficult, if not impossible, to analytically predict the equilibrium shapes of the membranes. Instead, specific shapes must be assumed and then the theory can assess which of those shapes will be energetically favorable. A more comprehensive theory would predict \textit{a priori} the shape of the membrane given parameters such as the depletant concentration and virus chirality.

    In this article, we take an important step towards developing such a comprehensive theoretical approach by carrying out Monte Carlo (MC) simulations of a discrete version of the %Jia et al.
    continuum model used by Jia et al.~\cite{jia2017chiral}.

    In our discrete model, the membrane is a triangular network consisting of hard spherical beads connected together by bonds~\cite{gompper1997network,gompper2004triangulated}. Fluidity of the membrane is imposed by allowing for bond reconnections~\cite{baumgartner1990crumpling}. We first determine the topology changes of nonchiral membranes, recapitulating the results of Ref.~\cite{boal1992topology} which include MC simulations showing a first order transition from a branched-polymer shape to a closed vesicle at low bending stiffness, as well as theoretical arguments indicating a transition  from flat disks to closed vesicles at higher bending stiffness. With greater computational power we are able to  extend the simulation results of Ref.~\cite{boal1992topology} to higher values of the membrane bending stiffness and study the transition from flat disks to closed vesicles.

    Then we consider the effects of chirality on the membrane shape, both in the interior and on the edge. Finally, inspired by the experiments of Refs.~\cite{gibaud2012reconfigurable,balchunas2019force}, where colloidal membranes were stretched using optical tweezers, we fix the locations of two beads on opposite sides of the membrane and measure the energy of the system as the distance between these two beads is varied. From this energy we can deduce the force needed to stretch the membrane and compare our results qualitatively to the experimentally measured values.

\section{Continuum model}
    The continuum model used by Jia et al.~\cite{jia2017chiral} is given by the following Hamiltonian, consisting of a bending term integrated over the area of the membrane, and an edge term integrated over the perimeter:
    \begin{equation}
        \mathcal{H} = \mathcal{H}_b + \mathcal{H}_e.
    \label{eq:Hamiltonian}
    \end{equation}
    The bending energy $\mathcal{H}_b$ is the Canham-Helfrich energy~\cite{canham1970minimum,helfrich1973elastic},
    \begin{equation}
    \mathcal{H}_b=\int \dd{A}\left[\frac{\kappa}{2}(2H)^2+\bar{\kappa}K \right],
    \end{equation}
    where $\kappa$ is the bending modulus, $H=(1/R_1+1/R_2)/2$ is the mean curvature, $\bar{\kappa}$ is the Gaussian curvature modulus, $K=1/(R_1R_2)$ is the Gaussian curvature, and $R_1$ and $R_2$ are the two principal radii of curvature of the surface.
    Using the Gauss-Bonnet theorem~\cite{struik1961lectures}, the integral of the Gaussian curvature over a surface with the topology of a disk can be rewritten as $\int \dd{A} K = 2\pi - \oint \dd{s} k_g$, where the last integral is over the edge of the surface. Ignoring a constant term, the bending energy $\mathcal{H}_b$ is therefore
    \begin{equation}
        \mathcal{H}_b= \int \dd{A}\frac{\kappa}{2}(2H)^2-\oint\dd{s}\bar{\kappa}k_g.
    \label{eq:bending_energy}
    \end{equation}
    The effective edge energy $\mathcal{H}_e$ proposed by Jia et al. is given by
    \begin{equation}
        \mathcal{H}_e= \oint \dd{s} \left[\lambda + \frac{B}{2}k^2+\frac{B'}{2}(\tau_g-\tau_g^*)^2\right],
    \label{eq:edge_energy}
    \end{equation}
    where $\lambda$ is the line tension, $B$ is the edge bending stiffness,  and $k$ is the curvature of the edge. The effect of chirality is introduced in the last term of the above equation with the edge torsional modulus $B'$ and the geodesic torsion $\tau_g=\vu{T}\cdot(\vu{n}_c\times\dv*{\vu{n}_c}{s})$. The geodesic torsion is the rate of rotation of the surface normal $\vu{n}_c$ around the tangent $\vu{T}$ of the edge~\cite{kleman,struik1961lectures}. The parameter $\tau_g^*$ is the spontaneous geodesic torsion of the edge and represents the chirality of the constituent virus particles which comprise the membrane. The sign of $\tau_g^*$ is determined by the chirality of the particles.

\section{Discrete model}
    We discretize the model of the previous section using a bond-and-bead model for self-avoiding membranes~\cite{gompper1997network}. The continuous two dimensional membrane surface is replaced by a triangular mesh $\mathcal{M}$ with hard-sphere beads of diameter $\sigma_0$ on vertices of the mesh which are connected by bonds. Each bead can be thought of as a coarse-grained group of virus particles.  A bond connecting two beads does not allow them to move farther apart than a distance $l_0$. For beads separated by a distance less than $l_0$ but greater than $\sigma_0$, we assume that there is no interaction between the beads, even when they are connected by bonds. We assume that the number of neighbors of any bead is between three and nine.

    The energy of a configuration of beads and bonds is given by discretizing Eqs.~(\ref{eq:bending_energy}) and (\ref{eq:edge_energy}), subject to the constraints imposed by the hard cores of the beads and the presence of bonds. For all but the last term in Eq.~(\ref{eq:edge_energy}), we use discretized forms of the terms appearing in the energy, Eq.~(\ref{eq:Hamiltonian}), that have appeared previously in the literature. The square of the discretized mean curvature $H(i)$ at bead $i$ is~\cite{itzykson1986,ItzyksonDrouffeV2,gompper1997network,espriu1987triangulated,meyer2003discrete}
    \begin{equation}
        [H(i)]^2 =\left[ \frac{1}{2\sigma_i} \sum_{j(i)}\frac{d_{ij}}{l_{ij}}(\va{r}_i-\va{r}_j)\right]^2,
    \label{eq:discrete-mean-curvature}
    \end{equation}
    where the sum is over the neighbors $j(i)$ of bead $i$.  The distance between beads $i$ and $j$ is $l_{ij}=|\va{r}_i-\va{r}_j|$ , where $\va{r}_i$ is the position vector of bead $i$. The length $d_{ij}$ is given by $d_{ij}=l_{ij}(\cot{\theta_1}+\cot{\theta_2})/2$, where $\theta_1$ and $\theta_2$ are the angles opposite bond $ij$ in the two triangles which meet at the bond, and $\sigma_i=\sum_{j(i)}d_{ij}l_{ij}/4$ is the area of the cell on the virtual dual lattice centered at bead $i$~\cite{itzykson1986,ItzyksonDrouffeV2}. Also, the discretized geodesic curvature can be written as~\cite{upadhyay2015gauss,mesmoudi2010discrete}
    \begin{equation}
        k_g(i) = \pi - \sum \theta_k,
    \label{eq:discrete-geodesic-curvature}
    \end{equation}
    where the sum is over the interior angles $\theta_k$ of the triangles meeting at bead $i$.

    Therefore, the discretized bending energy $E_b$ is
    \begin{equation}
        E_b = \frac{\kappa}{2}\sum_{i\in\mathcal{M}}\sigma_i (2H(i))^2 - \bar{\kappa}\sum_{i\in\partial \mathcal{M}}k_g(i) \dd{s}(i),
    \label{discrete-bending-energy}
    \end{equation}
    where $\partial\mathcal{M}$ is the edge of the triangular mesh $\mathcal{M}$, and the differential edge length is given by
    \begin{equation}
        \dd{s}(i) = \frac{1}{2}(l_{i,i-1}+l_{i,i+1}).
    \label{discrete-length}
    \end{equation}
    with $i-1, i+1$ denoting the neighboring beads of $i$ on the edge.

    Turning now to the edge energy, we write the discretized form $k(i)$ of the edge curvature as
    \begin{equation}
        k(i) = \frac{\theta_i}{\dd{s}(i)}
    \label{eq:discrete-curvature}
    \end{equation}
    where $\theta_i$ is the angle between bonds $i,(i-1)$ and $i,(i+1)$.

    To construct the discretized geodesic torsion, $\tau_g(i)$, we first define the discrete surface normal at bead $i$ \cite{Crane:2013:DGP},
    \begin{equation}
            \vu{n}_c(i) = \frac{\sum_j \theta_j \vu{n}_j}{|\sum_j \theta_j \vu{n}_j|},
    \label{eq:discrete-surface-normal}
    \end{equation}
    where the sum is over all of the  triangles with one vertex at $i$, $\vu{n}_j$ is the direction normal to the $j$th triangle, and $\theta_j$ is the interior angle of triangle $j$ at vertex $i$.

    To discretize the last term in Eq.~(\ref{eq:edge_energy}), we introduce a discretized geodesic torsion on bead $i$ by
    \begin{equation}
        \tau_g(i) = \frac{\va{r}_{i+1}-\va{r}_{i-1}}{l_{(i-1),(i+1)}}\cdot \left[\vu{n}_c(i)\times \frac{\vu{n}_c(i+1)-\vu{n}_c(i-1)}{2\dd{s}(i)}\right].
    \label{eq:discrete-geodesic-torsion}
    \end{equation}
    Thus, the discretized edge energy $E_e$ is given by
    \begin{equation}
        E_e = \sum_{i\in \partial \mathcal{M}}\dd{s}(i)\left\{\lambda+\frac{B}{2}k^2(i)+\left[\tau_g(i)-\tau_g^*\right]^2\right\},
    \label{discrete-edge-energy}
    \end{equation}
    and the total discretized energy is $E_M= E_b+E_e$.

\section{Monte Carlo Method}
\begin{figure}[t]
        \includegraphics[width=\linewidth]{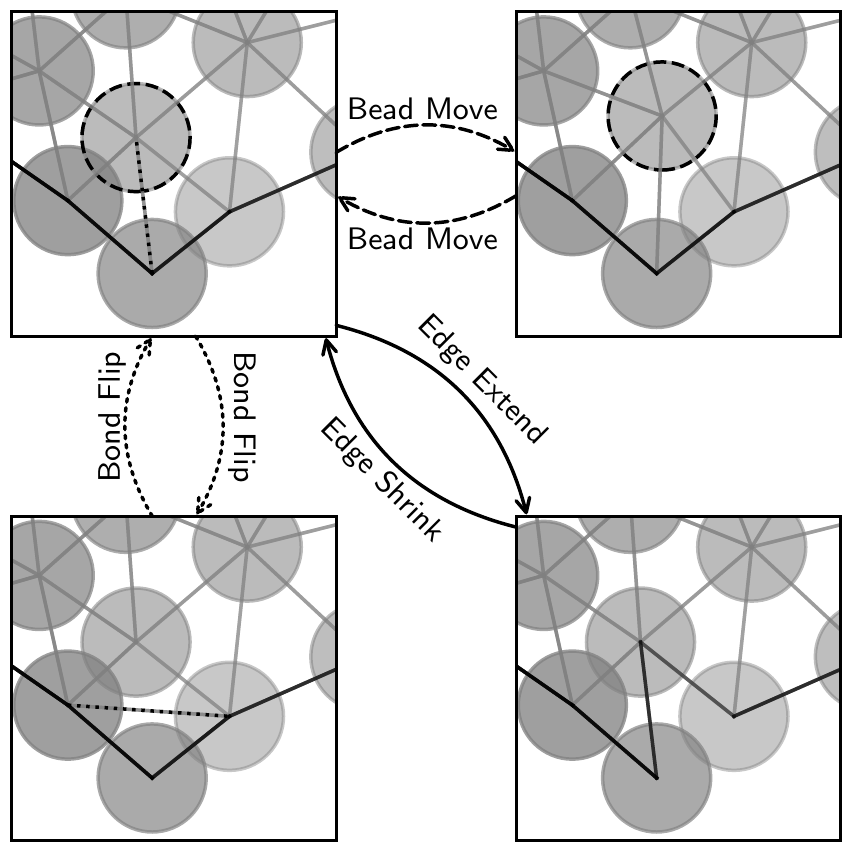}
        \caption{Monte Carlo updates for the beads, internal bonds, and edge bonds of the membrane. The updated bead and bond are highlighted with a black dashed line. Each update is reversible. The beads have a radius $\sigma_0$ and do not overlap due to hard-core repulsion. The apparent overlap in the figure is due to projection from three to two dimensions.}
        \label{fig:MCupdate}
    \end{figure}

    To sample the configuration space of the discrete model, we use Monte Carlo updates for the beads, the bonds and the edge as shown in Fig.~\ref{fig:MCupdate}. We measure lengths in units of $\sigma_0$ and energy in units of  $k_B T$. The bead positions are updated by choosing a bead at random and giving it a uniform random translation within a cube of side $2s=0.6$ centered on the bead. Bonds not on the edge are updated by choosing one at random and moving it as shown in Fig.~\ref{fig:MCupdate}~\cite{baumgartner1990crumpling}. Bonds on the edge are updated by replacing a single edge bond of a triangle bordering the edge with the two bonds of the same triangle as shown in  Fig.~\ref{fig:MCupdate}. Edge bonds can also be updated by reversing this process, replacing two neighboring edge bonds by one and creating a new triangle bordering the edge.
    The bead and bond update attempts are accepted with a probability specified by the Metropolis-Hasting algorithm~\cite{krauth2006statistical,hastings1970monte,metropolis1953equation}.

    In the simulation, $1.5\times10^5\times N$ MC steps are performed for each set of parameters, where $N$ is the number of particles. Each step is composed of $N$ attempts of moving a bead chosen at random, and $N$ attempts of flipping a bond chosen at random. We also make $\sqrt{N}$ attempts at edge shrinkage or extension. We equilibrate the system for the first $0.5\times10^5N$ steps, and then record the data every $N$ steps for the remaining $10^5N$ steps. The uncertainty in the observables is estimated using Sokal's method~\cite{sokal1997monte}. To avoid the formation of a hexatic phase, the maximum bond length is set to $l_0=1.68$ in all of our simulations~\cite{gompper2000melting}.

    \subsection{Disk to Vesicle Transition}
        \begin{figure}[t]
            \includegraphics[width=\linewidth]{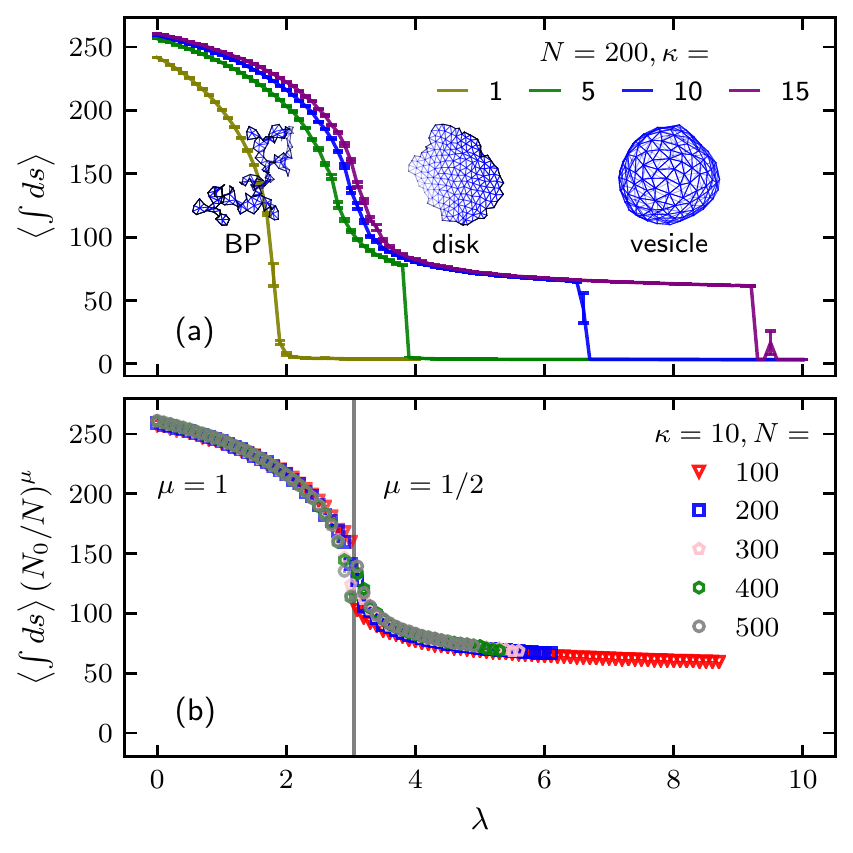}
            \caption{Simulation results for a membrane with bending stiffness $\kappa$ (in units of $k_\mathrm{B}T$) and edge tension $\lambda$ (in units of $k_\mathrm{B}T/\sigma_0$). All other moduli are zero.
            (a) Edge length $\left<\int \dd{s}\right>$ (measured in units of $\sigma_0$) as function of line tension $\lambda$ at different $\kappa$. The snapshots of the configurations have $\kappa=10$ and $\lambda=2.0$ (self-avoiding branched-polymer (BP) shape), $\lambda=4.5$ (disk), and $\lambda=7.0$ (vesicle).
            (b) Rescaled edge length vs. $\lambda$ for different system sizes $N$ for $\kappa=10$, showing the transition from the branched polymer shape to a disk shape. We used $N_0=200$ for the rescaling. Note that for each value of $N$, the rightmost data point is at the value of $\lambda$ at which the disk transitions to a vesicle.}
            \label{fig:O_scale_lam}
        \end{figure}

        We begin by investigating the simple topology change from a disk to a closed vesicle, which is driven by the competition between line tension and bending stiffness. For this simulation, we set the Gaussian curvature modulus to zero and include line tension as the only edge energy term. Due to advances in computing power over the past two decades, we are able to study membranes with larger bending stiffness (or equivalently, lower temperature) compared to previous work by Boal and Rao~\cite{boal1992topology}. With their lower value of bending stiffness, Boal and Rao studied the transition from a vesicle to a branched-polymer-like membrane with free edges, with the length of the perimeter scaling like the number of particles $N$. In our simulations, the bending stiffness is large enough that the state with free edges is a flat disk, with the length of the perimeter scaling like $N^{1/2}$.

        In Fig.~\ref{fig:O_scale_lam}a we plot the average membrane edge length $\left<\int \dd{s}\right>$ (measured in units of $\sigma_0$) as a function of line tension $\lambda$ for various values of $\kappa$, along with pictures of representative membrane configurations. The leftmost curve, corresponding to $\kappa=1$, shows a smooth transition from a branched-polymer shape directly to a closed vesicle. Since the perimeter in this case scales like $N$ (see supplemental material Fig.~\ref{fig:SI_O_lam}), our result is in quantitative agreement with that of Boal and Rao~\cite{boal1992topology}.  The other curves in  Fig.~\ref{fig:O_scale_lam}a, corresponding to higher values of the  bending stiffness $\kappa$, show a smooth transition from the branched polymer shape to a flat disk as the line tension increases, and then a sharp transition from the flat disk to a closed vesicle at higher line tension. Fig.~\ref{fig:O_scale_lam}b  shows how the perimeter scales with $N$ for the case of $\kappa=10$, and demonstrates the smooth transition from the branched polymer shape at lower edge tension $\lambda$ to the flat disk shape at higher $\lambda$.

        Though the critical line tension for the disk-vesicle transition is sensitive to both the bending stiffness $\kappa$ and the system size $N$, the value for $\lambda$ at the branch to disk transition barely changes as $\kappa$ or $N$ vary. In the our study of the effects of chirality in the next section, we take the bending stiffness $\kappa$ to be large enough that the membrane state with free edges is disk-like rather than branched-polymer-like.

        \begin{figure}[t]
            \includegraphics{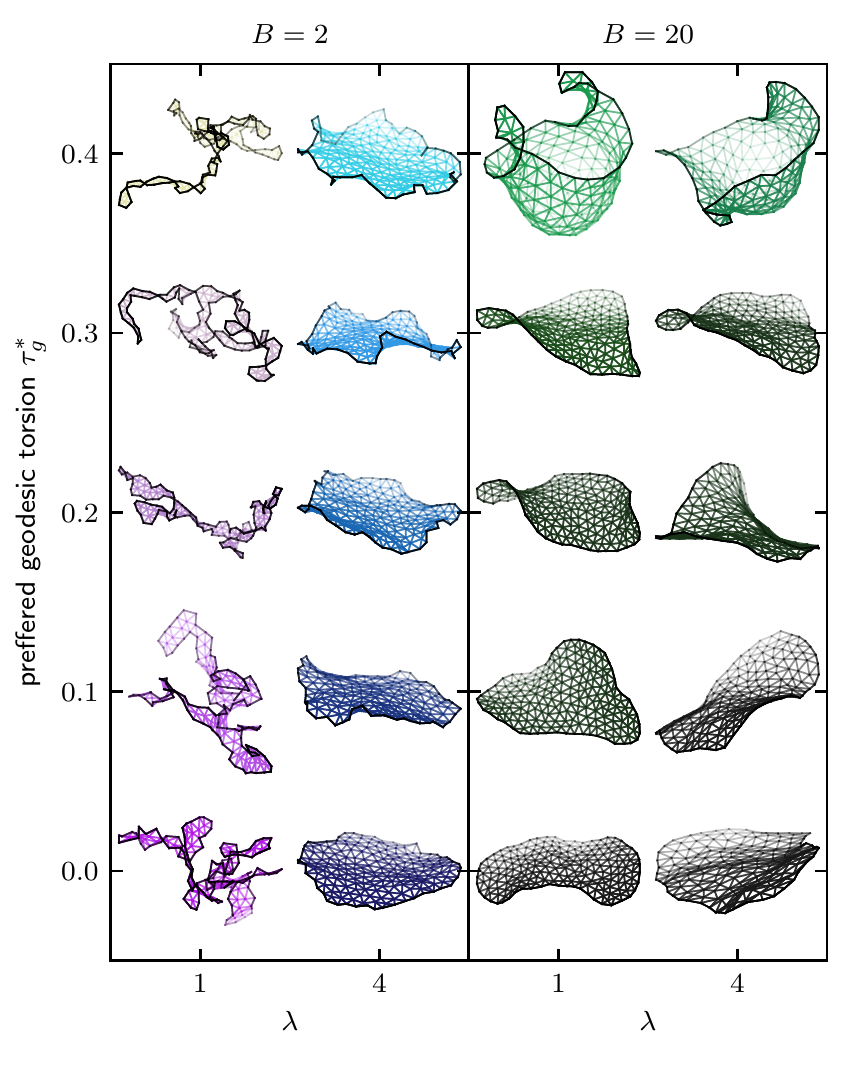}
            \caption{Snapshots of membranes with various values of the spontaneous edge geodesic torsion $\tau_g^*$ for two different line tensions $\lambda$;
            the membranes in the left panel have low edge bending stiffness, while those on the right have high edge bending stiffness ($B$ is measured in units of $k_BT\sigma_0$). The system size is $N=200$,the membrane bending stiffness is $\kappa=15$ (in units of $k_BT$) and the twist stiffness is $B'=100$ in units of $k_BT\sigma_0$.}
            \label{fig:config_pars}
        \end{figure}

    \subsection{Edge shape and fluctuation}
        Next, we add an edge bending stiffness, edge torsional stiffness, and a spontaneous geodesic torsion for the edge, but we disregard the Gaussian rigidity for simplicity. Even in the presence of a line tension $\lambda$, the edge of a membrane disk with no edge bending stiffness is jagged, as shown by the branched polymer and disk shapes in Fig.~\ref{fig:O_scale_lam}(a). Introducing a positive edge bending stiffness $B$ leads to a smoother edge and correlations between the tangent vectors along the edge. Fig.~\ref{fig:config_pars} shows this effect. In the left panel, where $B=2$, the membranes form branched-polymer shapes for $\lambda=1$, and disk-like shapes with rough edges for $\lambda=4$. As the spontaneous geodesic torsion $\tau_g^*$ of the edge increases, the branches form twisted ribbons. On the other hand, the disk shapes remain mostly flat as $\tau_g^*$ increases, but their edges exhibit localized regions of high twisting. These localized regions of twist lead to a rougher edge at the higher values of $\tau_g^*$. In the right panel of Fig.~\ref{fig:config_pars}, we see that the larger value of the edge bending stiffness leads to a smoother edge. Localized regions of edge twist are suppressed, but the high value of the twist stiffness $B'$ causes the entire membrane to warp like a saddle to allow the edge to twist to a degree that increases with increasing $\tau_g^*.$ Note that the high value of edge bending stiffness $B$ leads to disks with smooth edges even for low values of line tension such as $\lambda=1.$

\begin{figure}
            \includegraphics[width=\linewidth]{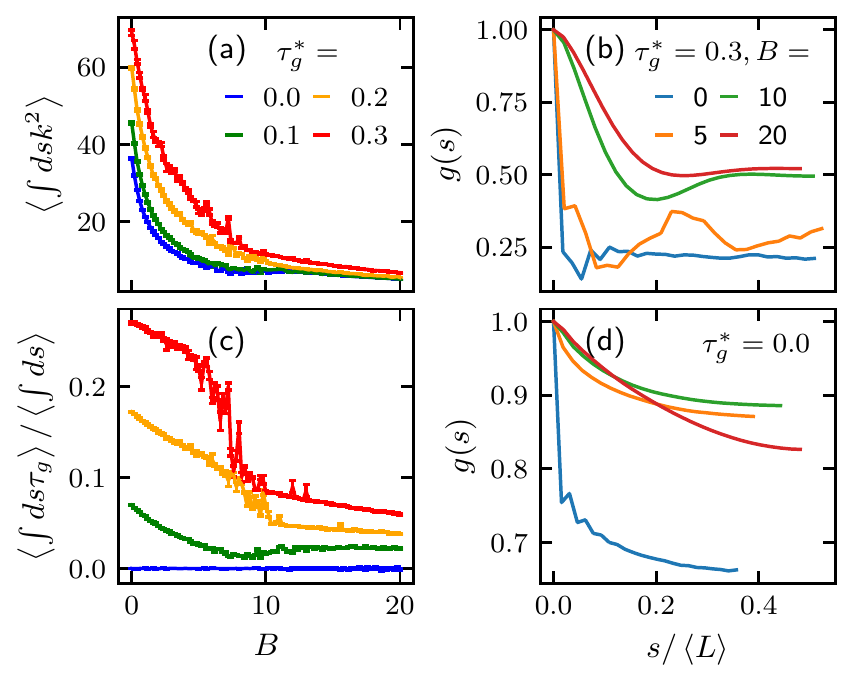}
            \caption{Geometrical properties of the membrane edge in the disk region of the phase diagram for a membrane with $N=200$, $\kappa=15$, $\lambda=4$ and strong torsional stiffness $B'=100$ (measured in units of $k_B T/\sigma_0$).
            (a) The average total square curvature (measured in units of $\sigma_0$) vs. $B$ (measured in units of $k_B T/\sigma_0$).
            (c) Average total geodesic torsion per average perimeter versus $B$.
            (b) The correlation function $g(s)=\left<\vu{n}(0)\cdot\vu{n}(s)\right>$ of the surface normal vector $\vu{n}$ along the edge versus neighbour distance $s$ divided by averaged number of beads on the edge $\left<L\right>$ for different values of $B$. The oscillation of $g(s)$ indicates twist of the edge.
            (d) $g(s)$ along the edge for the achiral case.}
            \label{fig:O_C_B}
        \end{figure}

        Next, we turn to a quantitative analysis of the membrane shapes. Fig.~\ref{fig:O_C_B}(a) shows that the total edge curvature squared decreases rapidly as the stiffness $B$ increases. A nonzero value of the spontaneous geodesic torsion of the edge causes the edge to twist, which necessarily leads to more edge curvature at small values of $B$. While we recognize that chirality, and more specifically, handedness, cannot be captured by a single pseudoscalar~\cite{harris_kamien_lubensky1999,efrati_irvine2014}, we quantify the handedness of the edge by dividing the average total geodesic torsion of the edge $\langle\int\mathrm{d}s\tau_g\rangle$ by the average perimeter $\langle\int\mathrm{d}s\rangle$ to associate an average rate of twist with the edge. Fig.~\ref{fig:O_C_B}(c) displays the average rate of twist of the edge vs. edge stiffness for various values of the spontaneous geodesic torsion $\tau_g^*$ and a large value of the twist modulus $B'$. When $\tau_g^*=0$, there is no preference for either handedness, and the average rate of twist vanishes. For nonzero $\tau_g^*$ and small $B$, the edge twists at rate that is close to the spontaneous geodesic torsion because $B'$ is so large that the cost for departure of $\tau_g$ from $\tau_g^*$ is high. But since twist of the edge requires curvature of the edge, the average twist decreases as $B$ increases.

        Fig.~\ref{fig:O_C_B}(b) and (d) show the correlation function $g(s)=\left<\vu{n}(0)\cdot\vu{n}(s)\right>$ of the surface normal vector $\vu{n}$ at the edge for the chiral (b) and achiral (d) cases, for various values of the edge bending stiffness $B$. When $B=0$, the correlation function decays rapidly since the edge is jagged. If $B=0$ but the membrane edge has a spontaneous geodesic torsion, the localized twist regions of the edge lead to less correlation than the achiral case. As $B$ increases, the correlation function for the chiral case starts to develop oscillations since the entire membrane is twisting like a potato chip.

        \begin{figure}[t]
            \includegraphics[width=\linewidth]{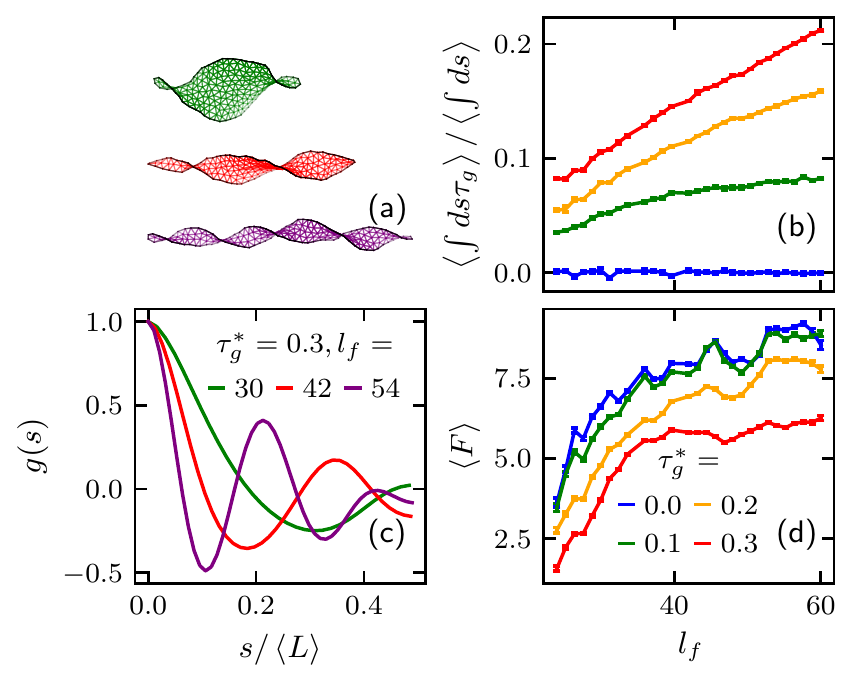}
            \caption{Results for a membrane under external forces. The parameters are such that the membrane forms a disk in the absence of external force: $N=200$, $\kappa=15$, $\lambda=4$, $B'=100$ and  $B=20$.
            (a) Snapshots of the membrane with $\tau_g^*=0.3$ and, from top to bottom, $l_f=30$, $42$ and $54$.
            (b) Average total geodesic torsion divided by average perimeter for $\tau_g^*=0$, $0.1$, $0.2$ and $0.3$, from bottom to top.
            (c) Correlation function $g(s)$ for the surface normal at the edge versus neighbour distance $s$ divided by averaged number of beads on the edge $\left<L\right>$.
            (d) Force required to impose the separation $l_f$, calculated as the derivative of the total average energy $\left<E\right>$ with respect to $l_f$. From top to bottom, the curves correspond to membranes with $\tau_g^*=0$, $\tau_g^*=0.1$, $\tau_g^*=0.2$ and $\tau_g^*=0.3$, from top to bottom.}
            \label{fig:O_C_L}
        \end{figure}

    \subsection{Ribbon formation under external force}
        Experiments show that a colloidal membrane disk subject to a stretching force by laser tweezers deforms into a twisted ribbon, with the twist increasing as the ends of the membrane are drawn apart~\cite{gibaud2012reconfigurable,balchunas2019force}. Motivated by this work, we fix the distance between two beads on the edge of our membrane, and find the shape as a function of the distance $l_f$ between these two beads. As the distance increases, the membrane forms a twisted ribbon, with the twist increasing with distance, as shown in Fig.~\ref{fig:O_C_L}(a). Note that a helicoid with right-handed helical edges has a positive geodesic torsion, in accord with the fact that we find right-handed ribbons when we pull on a membrane disk with positive $\tau_g^*$. If we reverse the sign of $\tau_g^*$, then we find that the handedness of the ribbons reverses (see supplemental material Fig.~\ref{fig:SI_O_L}). The average twist rate of the ribbon increases roughly linearly with extension $l_f$, except when $\tau_g^*=0$, in which case the membrane does not twist. (Recall that we have set $\bar\kappa=0$, so membrane bending energy does not give a tendency for the membrane to have negative Gaussian curvature).  Fig.~\ref{fig:O_C_L}(b) shows the correlation function $g(s)$ for the membrane normal at edge. The increase in oscillations with increasing values of $l_f$ correspond to an  increase in membrane twist with $l_f$. Finally, Fig.~\ref{fig:O_C_L}(d) shows the force required to hold the beads at separation $l_f$. The force is calculated by calculating the average energy as a function of $l_f$, and then differentiating with respect to $l_f$. The force rises linearly and then asymptotes to constant value. In the case of zero spontaneous geodesic torsion (uppermost curve), the force asymptotes to $2\lambda$. As $\tau_g^*$ increases, the value of the force plateau decreases. Similar results were found in a semi-analytic model which assumed the shape of the membrane is a helicoid~\cite{balchunas2019force}.

\section{Conclusion}
    Colloidal membranes take on a wide range of shapes beyond flat disks and closed vesicles due to their tendency to have free edges, and due to the chirality of their constituent particles. In this article, we determined the membrane shapes and their properties using Monte Carlo simulations with an effective energy that accounts for the liquid crystalline degrees of freedom near the edge using geometric properties of the edge. Our work extends semi-analytical approaches that make simplifying assumptions about the membrane shape~\cite{balchunas2019force}. The presence of the edges and the effective energy terms such as edge bending stiffness and edge torsional stiffness lead to a richer free energy landscape compared to existing studies of systems either with no edge~\cite{gompper2004triangulated,kohyama2003budding}, or with only line tension and bending stiffness~\cite{boal1992topology,zhao2005monte}. It would be natural to extend our work to consider more complex shapes such as membranes with the topology of a cylinder (two edges) or a trinoid (three edges), or even a M\"obius strip. The presence of free edges also suggests that we should study the effect of a nonzero Gaussian curvature modulus, which we disregarded here for simplicity. Finally, future work should test the validity of the assumptions of the effective theory by explicitly accounting for the liquid crystalline degrees of freedom in Monte Carlo simulations of the membrane.

\begin{acknowledgments}
This work was supported in part by the National Science Foundation through Grants No. CMMI-163552 and MRSEC-1420382.
\end{acknowledgments}

\bibliography{bibtex}

%%%%%%%%%% Merge with supplemental materials %%%%%%%%%%
\pagebreak
\widetext
\begin{center}
\textbf{\large Supplemental Materials: Shapes of fluid membranes with chiral edges}
\end{center}
%%%%%%%%%% Merge with supplemental materials %%%%%%%%%%
%%%%%%%%%% Prefix a "S" to all equations, figures, tables and reset the counter %%%%%%%%%%
\setcounter{equation}{0}
\setcounter{figure}{0}
\setcounter{table}{0}
\setcounter{page}{1}
\makeatletter
\renewcommand{\theequation}{S\arabic{equation}}
\renewcommand{\thefigure}{S\arabic{figure}}
\renewcommand{\bibnumfmt}[1]{[S#1]}
\renewcommand{\citenumfont}[1]{S#1}
%%%%%%%%%% Prefix a "S" to all equations, figures, tables and reset the counter %%%%%%%%%%

    \begin{figure}[!h]
    \centering
    \includegraphics[width=0.5\textwidth]{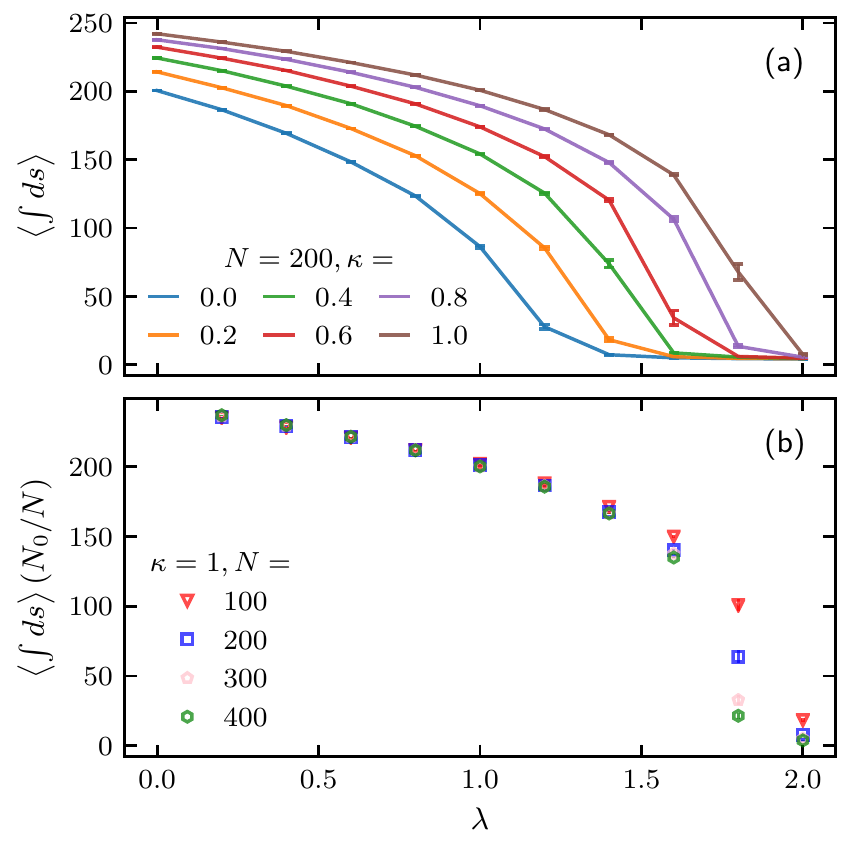}
    \caption{Simulations of a system with bending stiffness $\kappa$ and line tension $\lambda$ for the case of  small $\kappa$, where the membrane makes a continuous transition from the self-avoiding branched polymer phase to a vesicle.
    (a) Edge length $\left<\int \dd{s}\right>$ as a function of line tension $\lambda$ at different $\kappa$ ($\leq 1$).
    (b) Rescaled edge length vs. $\lambda$ for different system sizes $N$ and $\kappa=1$, showing the transition from branched polymer to vesicle. We used $N_0=200$ for the rescaling constant.
    }
    \label{fig:SI_O_lam}
    \end{figure}

    \begin{figure}[!h]
    \centering
    \includegraphics[width=0.5\textwidth]{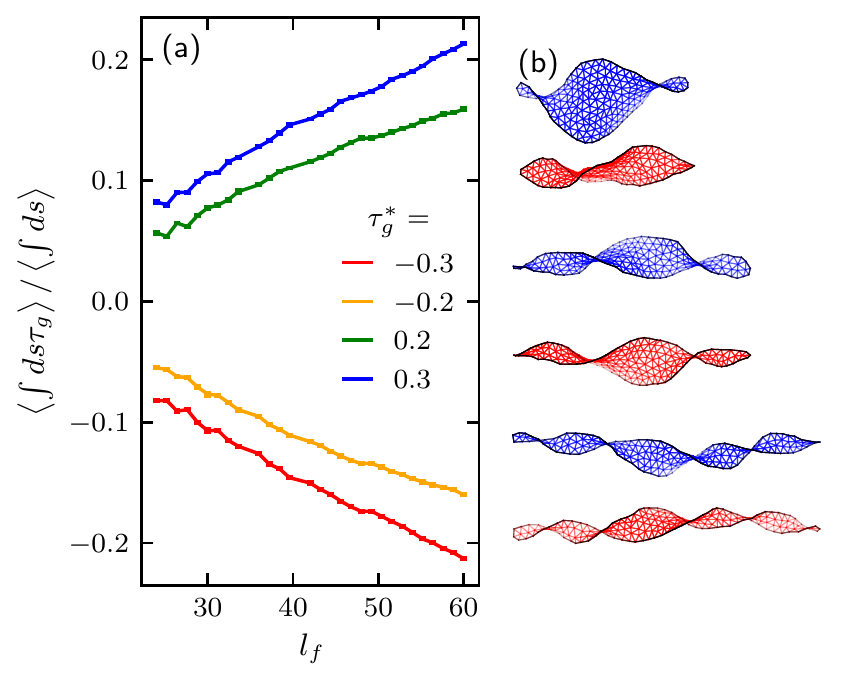}
    \caption{The effect of the sign of the spontaneous geodesic torsion of the edge on the shape of a membrane subject to an external force. (a) Average total geodesic torsion divided by average perimeter for $\tau_g^*=0.3,0.2,-0.2$ and $-0.3$, from top to bottom. (b) Snapshots of a membrane with $\tau_g^*=0.3$ (blue) and $\tau_g^*=-0.3$ (red), showing that membranes with opposite $\tau_g^*$ have opposite handedness, for $l_f = 30, 42$ and $54$ for the pairs from top to bottom.}
    \label{fig:SI_O_L}
    \end{figure}

\end{document}